\documentclass[11pt]{article}

\usepackage[preprint]{acl}

\usepackage{times}
\usepackage{latexsym}
\usepackage[T1]{fontenc}
\usepackage[utf8]{inputenc}
\usepackage{microtype}
\usepackage{inconsolata}

\usepackage{graphicx}
\usepackage{booktabs}
\usepackage{xcolor}
\usepackage{amsmath}
\usepackage{multirow}
\usepackage{subcaption}
\usepackage{xurl} 

\title{Progressive in Principle, Centrist in Practice: \\ LLM Political Bias Is Instrument-Dependent}

\author{Joel P. Barmettler \\
  Independent Researcher \\
  \texttt{jbarmettler@proton.me}}

\begin{document}
\maketitle

\begin{abstract}
Prior work establishes that instruction-tuned LLMs exhibit left-of-center political bias, but measures it exclusively through abstract questionnaires.
We show it does not predict how models vote on concrete policies.
We introduce a dual-instrument methodology grounded in Swiss direct democracy.
First, we administer the Smartvote questionnaire (75 policy questions) to 66 LLMs and compare their answers to those of 184 elected members of the Swiss National Council.
Second, we put 48 real federal referenda (Volksabstimmungen) to 9 flagship LLMs in four national languages and three information conditions, and compare their votes to the actual outcomes and to party recommendations (Parolen).

The instruments disagree.
(1)~The left-to-right agreement gradient that dominates Smartvote replicates prior work (mean $\rho = -0.77$).
On referenda it shifts to center-peaked: models align most with centrist Die~Mitte and FDP rather than leftist SP and Gr\"{u}ne (Wilcoxon $p = 0.008$).
(2)~For some models the \emph{language} of a question changes the answer: cross-linguistic consistency ranges from 50\% (Mistral) to 98\% (GPT-5.4).
(3)~Two models vote Nein on 83--94\% of referenda at similar rates on progressive and conservative proposals (binomial $p < 0.0001$), change-aversion rather than a left-right bias.
What prior work measured as ``leftward bias'' may not extend beyond abstract instruments: confronted with real decisions, LLMs behave less like coalition partners of the left than like cautious civil servants, centrist and inconsistent across languages.
\end{abstract}

\section{Introduction}

Four times a year, Swiss citizens vote on federal referenda.
For each proposal they receive an official booklet (the \emph{Abstimmungsb\"{u}chlein}) with a summary, background, arguments for and against, and the government's recommendation.
They then cast a binding vote.
The 48 federal votes held between March 2021 and March 2026 form a concrete record of political decision-making, with real proposals, real arguments, and real outcomes.
We gave the same booklets to nine large language models and asked them to vote.

The experiment probes a gap in work on LLM political bias.
A substantial body of work establishes that instruction-tuned LLMs lean left of center on abstract political instruments: Political Compass Tests, opinion surveys, and voting advice applications (VAAs) such as Germany's Wahl-O-Mat or Switzerland's Smartvote~\citep{hartmann2023political, rozado2023political, rozado2024political, rutinowski2024self, rozado2025measuring, motoki2025assessing}.
The finding replicates across dozens of models and national contexts, yet every instrument shares one property: it is abstract.
VAAs ask about policy \emph{proposals} and compass tests about \emph{principles}; neither measures political \emph{behavior}: how a model answers when given a concrete policy with its tradeoffs and arguments from both sides.
If models answer differently when the same politics arrives as a concrete proposal rather than an abstract statement, the established leftward bias may not describe how they act on real decisions.

Switzerland is uniquely suited to this test.
Its 200-member National Council spans six major parties on a well-defined left-right axis (SP, Gr\"{u}ne, GLP, Die Mitte, FDP, SVP)~\citep{linder2021swiss}, finer resolution than the binary US framing of prior work.
184 elected members completed the 2023 Smartvote VAA, which makes abstract measurement comparable to earlier studies.
Its direct-democratic tradition supplies a second, independent instrument: concrete referendum decisions accompanied by multilingual government booklets and binding party recommendations (\emph{Parolen}).
Its four national languages (German, French, Italian, Romansh) turn query language into a natural experiment.

We administer both instruments to the same models and ask three questions.
\textbf{RQ1:} do abstract questionnaires predict how LLMs behave on concrete policy decisions?
\textbf{RQ2:} does the language of a political question change the answer?
\textbf{RQ3:} do LLM positions match how the electorate actually voted?

The two instruments give different answers.
On Smartvote we replicate prior work: 66 models from 27 families converge left of the parliamentary centroid and show the expected left-to-right agreement gradient.
On the 48 referenda that gradient does not attenuate but moves from left-peaked to center-peaked, and a model's Smartvote position does not predict its referendum direction ($\rho = -0.25$, $p = 0.55$).
Two further results compound this decoupling: cross-linguistic consistency ranges from 50\% to 98\%, and two models vote Nein on 83--94\% of referenda regardless of political direction.

We frame these results through the \emph{invisible coalition partner}.
In Swiss consensus democracy, policy emerges from negotiation among governing parties; an LLM consulted on political questions acts as an unelected participant whose answers shape the information environment.
This partner's measured positions depend on the instrument used and, for some models, on the language of the question.

\section{Background and Related Work}

\subsection{The Swiss Political and Linguistic Context}

Three features of Swiss politics structure our analysis.
Six major parties span a well-defined left-right axis~\citep{linder2021swiss}, which gives finer resolution than the binary US framing and separates convergence from measurement noise.
The Smartvote VAA (run since 2003, 2.1M recommendations in 2023) provides standardized 75-question profiles for every elected member~\citep{smartvote2023}.
The referendum system yields concrete decisions with real outcomes, an instrument that measures how an agent \emph{votes} under real tradeoffs rather than what it \emph{states} in the abstract.
The four national languages create a natural experiment structured by the \emph{R\"{o}stigraben}, the documented German--French voting divide on EU integration, welfare, and military policy~\citep{linder2021swiss}; Romansh ($\sim$60{,}000 regular speakers) additionally stress-tests low-resource handling.

\subsection{Political Bias in Large Language Models}

Instruction-tuned LLMs lean left of center on abstract political instruments.
First shown by \citet{liu2022quantifying} for GPT-2, the result has been replicated across dozens of models and national contexts: pro-environmental, left-libertarian orientations robust across languages~\citep{hartmann2023political}; leftward positioning across 15 orientation tests, 24 models, and four measurement paradigms~\citep{rozado2023political, rozado2024political, rozado2025measuring}; and the same pattern on G7 and US/Brazilian/UK questionnaires~\citep{rutinowski2024self, motoki2024more}.

The bias is neither uniform across issues nor stable under pressure.
\citet{ceron2024beyond} found left-leaning outputs on environment and welfare but right-leaning ones on law and order, and under argumentative pressure models hold their original stance on up to 95\% of topics~\citep{press2025framework}.
RLHF is the likely mechanism: it shifts opinions toward liberal, higher-income demographics~\citep{santurkar2023whose}, and instruction tuning moves base models leftward from a near-center start~\citep{faulborn2025only}.
Robustness is contested too: PCT results behave as a ``spinning arrow'' under prompt variation~\citep{rottger2024political} and exaggerate bias relative to theory-grounded measures~\citep{faulborn2025only, fujimoto2023revisiting}, though model identity explains over 90\% of variance versus under 2\% for phrasing, keeping relative comparisons robust~\citep{sakhawat2026political}.

Every study reviewed above nonetheless uses abstract instruments; none tests whether the measured bias predicts behavior on concrete policy decisions with real tradeoffs and arguments.

\subsection{From Bias to Convergence}

Models also converge on the \emph{same} position.
Reported homogeneity is high~\citep{rozado2024political}: low-variance similarity on the Wahl-O-Mat~\citep{rettenberger2025assessing}, 96.3\% of models in the libertarian-left PCT quadrant~\citep{sakhawat2026political}, and $>$75\% alignment with left-wing parties on Germany's 2025 Wahl-O-Mat~\citep{dormuth2026cautionary}; neither model scale nor openness predicts positioning~\citep{peng2024beyond}.
One counterclaim holds that models ``reflect the ideology of their creators''~\citep{buyl2025llms, pnasnexus2025censorship}: Chinese models favor centralized governance, Western models liberal-democratic values.
We test it by placing Chinese and Western models in one political space.

\subsection{Language and Refusal}

Language adds a further axis of instability.
The same model gives systematically different responses across languages~\citep{nadeem2026bias}, query language can flip ideological framing (Russian vs.\ Ukrainian; \citealp{languageyouask2026}), and translating Wahl-O-Mat items from German to English shifts positioning by 3.5 points~\citep{exler2025large}.
Refusal compounds this: it is language-dependent (Bard refused 90\% of Putin queries in Russian vs.\ 19\% in English; \citealp{urman2024silence}) and reached 79\% in Gemini during the 2024 EU elections~\citep{haman2024chatbots}.
We test both effects within a single multilingual country, using identical questions in four national languages.

No prior work combines an abstract and a concrete instrument to test whether abstract bias predicts concrete behavior, or whether questionnaire convergence predicts convergence on real policy decisions.

\section{Methodology}

Our design uses two independent instruments grounded in Swiss democratic reality, one abstract (Smartvote) and one concrete (Volksabstimmungen), administered to the same set of LLMs.
This dual-instrument approach enables both replication of prior findings and a novel test of whether abstract bias predicts concrete behavior.

\subsection{Instrument 1: Smartvote Questionnaire}

The Smartvote questionnaire from the 2023 Swiss National Council election comprises 75 questions across 14 policy categories~\citep{smartvote2023}.
Each question presents a policy proposal (e.g., ``Should Switzerland introduce a national minimum wage of CHF~4,000 per month?'') with pro and contra arguments.
Respondents answer on a four-point scale: \emph{Ja} (yes, 100), \emph{Eher Ja} (rather yes, 75), \emph{Eher Nein} (rather no, 25), \emph{Nein} (no, 0).
We exclude 8 budget-allocation questions whose distribute-a-budget format is incompatible with PCA, leaving 67 policy-stance questions; this removes some fiscal-policy items where left-right differences are often sharpest.

Our benchmark is the 184 members from the six major parties who completed Smartvote and won seats in October 2023 (Smartvote GraphQL API); 194 of 200 completed it, 10 from minor parties.
They belong to the six major parties: SVP ($n = 58$), SP ($n = 41$), Die Mitte ($n = 29$), FDP ($n = 25$), Gr\"{u}ne ($n = 21$), GLP ($n = 10$).

\subsection{Instrument 2: Volksabstimmungen}

We scraped 50 federal referenda held between March 2021 and March 2026 from the Zurich statistics API.
Two votes with incompatible binary formats (a Stichfrage and a Direkter Gegenentwurf, where Ja/Nein does not map to support/opposition) were excluded, leaving 48 votes.
For each vote, we obtained: the official multilingual information texts in all four national languages (German, French, Italian, Romansh), national and cantonal voting results, and party Parolen (voting recommendations) from six parties.

Parolen (official party recommendations: \emph{Ja}, \emph{Nein}, \emph{Stimmfreigabe} (free vote), or \emph{keine Angabe}) are the political benchmark here, enabling a party-level rather than individual-level comparison; we analyze only directional (Ja/Nein) Parolen.

The 48 votes span taxation, healthcare, immigration, energy, civil liberties, agriculture, and institutional reform (e.g., the Verh\"{u}llungs\-verbot, Klima\-schutz\-gesetz, and E-ID-Gesetz).
The winning share ranges from 50.4\% to 84.1\%, providing variation in political contestedness.

\subsubsection{Cross-linguistic design.}
Each vote was presented in four languages using the official government texts.
This design tests whether positions are language-invariant or whether query language induces systematic shifts, an ``artificial R\"{o}stigraben.''

\subsubsection{Information conditions.}
Each vote was presented under three levels of detail:
\begin{enumerate}
    \item \textbf{Brief} (\emph{In K\"{u}rze}): Summary paragraph only.
    This is the primary analysis condition because it most closely parallels the abstract framing of Smartvote questions, enabling a controlled comparison between instruments while minimizing confounds from argument exposure.
    \item \textbf{Detailed} (\emph{In K\"{u}rze + Im Detail}): Summary plus factual background.
    \item \textbf{Full text}: All chapters including pro and contra arguments.
\end{enumerate}
The three conditions test whether added context, particularly both sides' arguments, shifts model positions.

\subsection{Model Selection}

For the Smartvote experiment, we queried 72 LLMs and analyze the 66 that met the refusal and completeness criteria (\S\ref{sec:refusal}), from 27 model families across five countries (USA, China, France, Canada, Israel), selected to maximize diversity along provider nationality, licensing (26 open-source, 40 closed-source), capability tier, and architecture.
Data was collected in multiple rounds between January 2025 and March 2026 to capture models as they were released; the analysis pools all models.

For the Volksabstimmungen experiment, we selected 9 flagship models representing the frontier of each major provider family (Table~\ref{tab:flagships}): four countries, two open-source, seven closed-source.
Gemini~3.1~Pro was excluded from Volksabstimmungen analysis due to a 98\% refusal rate in the primary condition (German, brief), but its refusal patterns across languages and conditions are analyzed separately (\S\ref{sec:refusal}).

\begin{table}[t]
\centering
\caption{Nine flagship models used in the Volksabstimmungen experiment.}
\label{tab:flagships}
\small
\begin{tabular}{llll}
\toprule
Model & Provider & Country & License \\
\midrule
GPT-5.4 & OpenAI & USA & Closed \\
Claude Opus 4.6 & Anthropic & USA & Closed \\
Gemini 3.1 Pro & Google & USA & Closed \\
DeepSeek V3.2 & DeepSeek & China & Open \\
Llama 4 Maverick & Meta & USA & Open \\
Grok 4.20 & xAI & USA & Closed \\
Mistral Large 2512 & Mistral & France & Closed \\
Qwen 3.5 Plus & Alibaba & China & Closed \\
Command A & Cohere & Canada & Closed \\
\bottomrule
\end{tabular}
\end{table}

\subsection{Data Collection}

All models were queried via the OpenRouter API with deterministic parameters: \texttt{temperature=0.0}, \texttt{seed=42}.

\textbf{Smartvote.}
Each query included the question text, supplementary information, pro and contra arguments, and any glossary items provided by Smartvote, all in German.
A system prompt instructed the model to respond with exactly one of four options (\emph{Ja}/\emph{Eher Ja}/\emph{Eher Nein}/\emph{Nein}) and no elaboration.
Unparseable responses were coded as $-1$ (missing) and imputed with the neutral midpoint (50) in the PCA and agreement vectors.

\textbf{Volksabstimmungen.}
Each query included the vote title, date, relevant text chapters (per condition), and a language-appropriate system prompt instructing the model to respond with exactly \emph{Ja} or \emph{Nein} (or the equivalent in French, Italian, or Romansh) and no elaboration.
Exact prompt templates are available in the code repository.
The design is 9~models $\times$ 48~votes $\times$ 4~languages $\times$ 3~conditions $= 5{,}184$ queries; 90 booklets lacked the required chapter, leaving 5{,}094 responses.
Responses were parsed via language-specific keyword matching; ambiguous or multi-paragraph responses were coded as refused ($-1$).

\subsection{Statistical Approach}

We fit PCA on the $184 \times 67$ matrix of parliamentarian answer vectors~\citep{pedregosa2011scikit} following \citet{hartmann2023political}, defining a political space from parliamentarians alone and projecting LLM vectors into it post-hoc.
PC1 is negated in all figures so left-wing parties appear on the left (display convention; PCA is sign-invariant).
We validate it against the known party ordering (Spearman) and silhouette score.

\begin{figure}[t]
    \centering
    \includegraphics[width=\columnwidth]{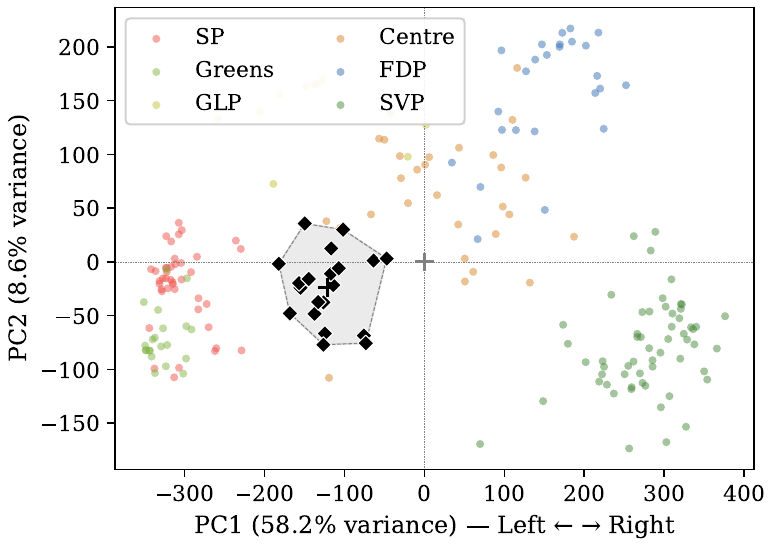}
    \caption{Smartvote: 2D PCA projection of 184 Swiss parliamentarians (colored by party) and flagship LLMs (black diamonds). All LLMs cluster in the center-left, nearest to GLP and Die Mitte, displaced from the parliamentary centroid~(+).}
    \label{fig:pca2d}
\end{figure}

\begin{figure}[t]
    \centering
    \includegraphics[width=\columnwidth]{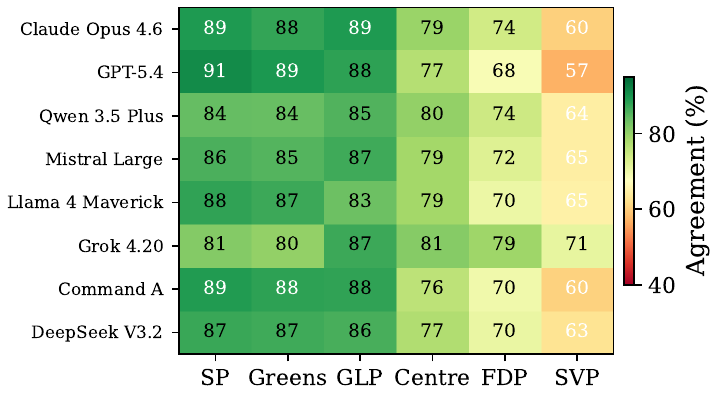}
    \caption{Smartvote agreement heatmap (8 flagship LLMs $\times$ 6 parties, left to right). A general gradient from high agreement (left parties) to low agreement (right parties) is visible across most model rows, though not all are strictly monotonic.}
    \label{fig:heatmap}
\end{figure}

Smartvote LLM--party agreement is a squared-difference score:
\begin{equation}
\label{eq:agreement}
    \text{Agreement} = 100 \left(1 - \frac{\overline{(a_{\text{LLM}} - a_{\text{party}})^2}}{100^2}\right)
\end{equation}
The overline denotes the mean over shared questions across a party's members, so the score is normalized to $[0, 100]$.
A mean-absolute-difference variant preserves each model's left-to-right agreement ordering ($\rho = 0.984$), so the gradients reported below are not quadratic-penalty artifacts (Appendix~\ref{app:stats}).
Volksabstimmungen agreement is the percentage of directional-Parole votes where the model's Ja/Nein matches the party.

Each research question uses one primary test.
RQ1: within each model and instrument we correlate party position (left$\to$right) with agreement (Spearman $\rho$; negative $=$ left-peaked), then compare the 8 paired $\rho$ across instruments with a Wilcoxon signed-rank test.
RQ2: per-model four-language consistency, pairwise McNemar tests (BH-corrected), and a Spearman correlation of cantonal DE--FR voting gaps against model DE--FR answer gaps.
RQ3: per-model popular-vote alignment (chi-square for heterogeneity), binomial tests for Ja/Nein tendency, and a direction-conditional (progressive- vs.\ conservative-Ja) split separating ideology from status-quo preference.
We apply Benjamini--Hochberg correction~\citep{benjamini1995controlling} across the 30 primary tests, and separately within the McNemar and per-category families (Appendix~\ref{app:stats}).
Permutation tests, 95\% bootstrap CIs~\citep{efron1993bootstrap}, and imputation-sensitivity checks are detailed in Appendix~\ref{app:stats}.

\section{Results}

We present results in four subsections.
Section~\ref{sec:smartvote} establishes the Smartvote baseline (replicating prior findings).
Section~\ref{sec:volksabstimmungen} presents the Volksabstimmungen results and tests whether abstract bias predicts concrete behavior (RQ1).
Section~\ref{sec:roestigraben} examines cross-linguistic consistency (RQ2).
Section~\ref{sec:refusal} analyzes refusal patterns across languages and conditions.

\subsection{Smartvote: Convergent Center-Left Positioning}
\label{sec:smartvote}

The first two principal components explain 66.8\% of parliamentary variance; PC1 aligns with the established left-right ordering (Spearman $\rho = -0.943$, $p = 0.005$) and the silhouette score (0.398) confirms party separation.
All 66 LLMs cluster in one narrow region, displaced from the parliamentary centroid by 131.2 in 2D (Figure~\ref{fig:pca2d}; permutation $p = 0.0002$; the PC1 displacement is 3.64 between-model SDs).
This holds under all imputation values (all $p = 0.0002$).
No structural variable predicts positioning: geographic origin ($p = 0.872$), open- vs.\ closed-source ($p = 0.870$), and temporal drift (12 pairs, 6 each way, $p = 1.000$) are all null.
The LLM centroid sits nearest GLP and Die~Mitte; individual models span SP to Die~Mitte.

The Smartvote agreement heatmap (Figure~\ref{fig:heatmap}) shows the expected left-to-right gradient: highest agreement with SP and Gr\"{u}ne, lowest with SVP, though 37 of 66 models deviate from strict monotonicity, most commonly by peaking at GLP.
This gradient is the signature of the abstract instrument and the pattern against which we test the referendum results.

\subsection{LLMs in the Volksabstimmung}
\label{sec:volksabstimmungen}

\subsubsection{The Gradient Flips (RQ1).}
\begin{figure}[t]
    \centering
    \includegraphics[width=\columnwidth]{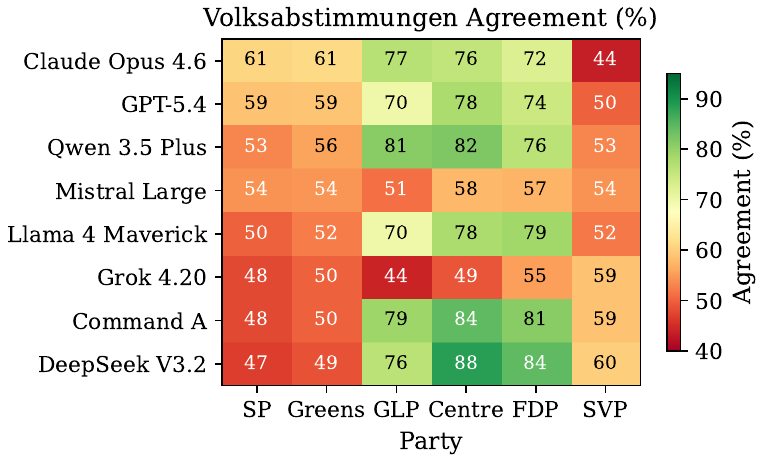}
    \caption{Volksabstimmungen Parolen agreement heatmap (8 flagship LLMs $\times$ 6 parties). Compare to Figure~\ref{fig:heatmap}: the left-peaked gradient has shifted to center-peaked. Most models now agree most with Die Mitte and FDP, not with SP and Gr\"{u}ne; Claude is the exception, peaking at GLP on both instruments.}
    \label{fig:volks-heatmap}
\end{figure}

The left-to-right gradient that dominates Smartvote shifts from left-peaked to center-peaked on Volksabstimmungen (Figure~\ref{fig:volks-heatmap}).
Across the 8 flagship models, the Spearman correlation between party position (left$\to$right) and agreement reverses from negative on Smartvote (mean $\rho = -0.77$, 6/8 significant) to positive or flat on the referenda (mean $\rho = +0.34$, 0/8 significant).
A Wilcoxon signed-rank test on the 8 paired $\rho$ gives $p = 0.008$ (BH-adjusted $0.017$), the smallest value attainable at $n = 8$.

\begin{figure}[t]
    \centering
    \includegraphics[width=\columnwidth]{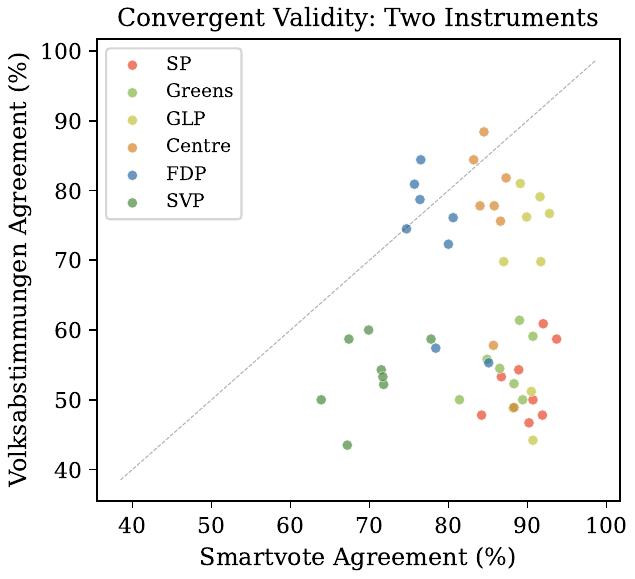}
    \caption{Convergent validity: Smartvote agreement vs.\ Volksabstimmungen agreement per model--party pair. If abstract and concrete instruments measured the same construct, points would cluster along the diagonal. Instead, left-party pairs (SP, Gr\"{u}ne; warm-colored markers in the upper-left region) fall far below the diagonal, indicating that high abstract agreement does not translate to high concrete agreement. Center-right pairs (Die Mitte, FDP) remain closer to the diagonal.}
    \label{fig:convergent-validity}
\end{figure}

The flip is a selective collapse rather than a mirror.
Agreement falls for every party but far more for the left (SP $-37$pp, Gr\"{u}ne $-33$pp) than the center-right (FDP $-6$pp), moving the peak from SP/Gr\"{u}ne to Die~Mitte in 5 of 8 models; Claude alone still peaks at GLP.
These per-party magnitudes partly reflect the 4-point-to-binary scale change and should not be over-read, but the direction change does not depend on them.
Concretely, models prefer SP over Die~Mitte by $+4$pp on Smartvote yet prefer Die~Mitte by $+22$pp on Volksabstimmungen (Wilcoxon $p = 0.016$; Figure~\ref{fig:instrument-shift}).
And a model's Smartvote PC1 position does not predict its referendum direction (SP-minus-SVP agreement; $\rho = -0.25$, $p = 0.55$): abstract and concrete positioning are decoupled (Figure~\ref{fig:convergent-validity}).

\subsubsection{Popular Vote Alignment (RQ3).}
Alignment with the popular vote outcome ranges from 97.9\% (GPT-5.4) to 60.4\% (Grok), a significant spread ($\chi^2 = 35.07$, $p < 0.0001$; Table~\ref{tab:popular-alignment}).

\begin{table}[t]
\centering
\caption{Popular vote alignment by model (German, brief condition). Close $=$ winning share $\leq$55\%; decisive $=$ winning share $>$60\%. $n = 48$ except DeepSeek (46) and Qwen (47), which refused some votes.}
\label{tab:popular-alignment}
\small
\setlength{\tabcolsep}{5pt}
\begin{tabular}{lrrr}
\toprule
Model & Overall & Close & Decisive \\
\midrule
GPT-5.4 & 97.9\% & 92.3\% & 100\% \\
Claude Opus 4.6 & 95.8\% & 92.3\% & 96.3\% \\
Qwen 3.5 Plus & 85.1\% & 69.2\% & 96.2\% \\
Command A & 79.2\% & 53.8\% & 88.9\% \\
Llama 4 Maverick & 72.9\% & 61.5\% & 70.4\% \\
Mistral Large & 70.8\% & 53.8\% & 85.2\% \\
DeepSeek V3.2 & 69.6\% & 38.5\% & 88.0\% \\
Grok 4.20 & 60.4\% & 53.8\% & 74.1\% \\
\bottomrule
\end{tabular}
\end{table}

\subsubsection{Systematic Nein Tendency.}
Two models show extreme Nein bias: Grok votes Ja on only 6.2\% of referenda (3/48, binomial $p < 0.0001$) and Mistral on 16.7\% (8/48, $p < 0.0001$); no other model deviates significantly from 50\%.
The Nein bias is change-aversion, not ideology.
Of the 48 votes, 23 are progressive-Ja and 17 conservative-Ja; Grok votes Nein on 91\% of the former and 94\% of the latter, Mistral on 78\% and 88\%, with no significant difference (Fisher's $p = 1.0$, $0.68$; Figure~\ref{fig:nein-tendency}).

\subsubsection{Secondary Patterns.}
Three secondary patterns round out the picture.
Models agree with the Bundesrat recommendation on 48--89\% of votes, but for centrist models that rate is indistinguishable from their Die~Mitte agreement (DeepSeek 89.1\% vs.\ 88.4\%), reflecting centrism rather than deference to authority.
Models are also more similar to one another than politicians within a single party are (mean pairwise Smartvote agreement among the 19 family flagships 92.3\% vs.\ 88.6\% within party; permutation $p < 0.001$), suggesting optimization compresses political variance beyond shared partisanship.
Finally, models are 81--96\% consistent across the three information conditions and rarely abstain even on the 10 votes where at least one major party issued no directional Parole (3 of them with two or more); added arguments do not systematically move positions, except Grok, whose Ja rate rises from 6\% to 25\% under full text (exploratory, $n = 7$ flips among the 36 votes it answered).

\subsection{The R\"{o}stigraben in Silicon (RQ2)}
\label{sec:roestigraben}

\subsubsection{Cross-Linguistic Consistency.}
Cross-linguistic consistency varies sharply (Table~\ref{tab:language}): GPT-5.4 gives the same answer in all four languages on 97.9\% of votes, Mistral on only 50.0\% and Llama 58.3\%.

\begin{table}[t]
\centering
\caption{Cross-linguistic consistency (proportion of votes on which every parseable answer across the four languages agrees; refused answers are dropped from the comparison and analyzed separately in \S\ref{sec:refusal}) and Ja rate by language (in\_k\"{u}rze condition).}
\label{tab:language}
\small
\begin{tabular}{lrrrrr}
\toprule
Model & Consist. & Ja\textsubscript{de} & Ja\textsubscript{fr} & Ja\textsubscript{it} & Ja\textsubscript{rm} \\
\midrule
GPT-5.4 & 97.9\% & 48\% & 48\% & 46\% & 46\% \\
Claude & 93.8\% & 50\% & 46\% & 50\% & 48\% \\
Qwen & 83.3\% & 45\% & 52\% & 51\% & 46\% \\
Grok & 83.3\% & 6\% & 17\% & 10\% & 13\% \\
DeepSeek & 81.2\% & 54\% & 60\% & 56\% & 58\% \\
Command A & 77.1\% & 50\% & 50\% & 40\% & 54\% \\
Llama & 58.3\% & 65\% & 54\% & 94\% & 58\% \\
Mistral & 50.0\% & 17\% & 19\% & 38\% & 82\% \\
\bottomrule
\end{tabular}
\end{table}

McNemar tests, BH-corrected within the family of 48 pairwise tests, find significant language effects only for Llama (3 pairs, all involving Italian) and Mistral (4 pairs).
Mistral is the extreme case: its Ja rate swings from 17\% in German to 82\% in Romansh, though 42\% of Romansh queries are refused (\S\ref{sec:refusal}), so the 82\% rests on approximately 28 responses and would fall to $\sim$48\% if refusals were Nein.
Even so, the same model yields a change-averse Nein-voter in German and a Ja-voter in Romansh on identical questions.

\subsubsection{The R\"{o}stigraben.}
These language shifts do not track the actual R\"{o}stigraben: cantonal DE--FR voting gaps and model DE--FR answer gaps do not correlate ($\rho = -0.231$, $p = 0.114$), though at $n = 48$ power to detect $\rho = 0.3$ is only $\sim$55\%.
No individual model is significant (Appendix~\ref{app:stats}).
The language sensitivity is real but reflects model-internal processing differences, not the Swiss linguistic divide (Appendix~\ref{app:figures}).

\subsection{Refusal Patterns}
\label{sec:refusal}

Refusal adds a further asymmetry: some models give no answer at all, and whether they do depends on language and context.
On Smartvote we excluded four models refusing over 25\% (worst: Gemini~3.1~Pro, 83.6\%), one that returned 8 of 75 answers, and one duplicate, leaving 66 of 72 (Appendix~\ref{app:stats}).
On Volksabstimmungen, refusal depends on language and context across three providers, not on model identity alone.
Gemini refuses 98\% of brief German votes but 13\% with full context and 54\% in Romansh; Grok, 50--73\% in French and 22--27\% in German under the two longer conditions but under 9\% elsewhere; Mistral, 42\% in brief Romansh (inflating its Romansh Ja rate, \S\ref{sec:roestigraben}).
That safety filters can be bypassed by adding context or switching language means minority-language speakers may face differential access to answers, not merely different ones (per-condition rates in Appendix~\ref{app:figures}).

\section{Discussion}

\subsection{An Instrument-Dependent Political Character}

Our dual-instrument design shows that a model's measured position depends on the instrument.
On Smartvote they behave like center-left progressives: they agree most with SP and Gr\"{u}ne and sit significantly left in 9 of 13 policy categories, the four nulls being democracy and media, security, foreign policy, and ethics.
Their convergence is so tight that the spread of 66 models on PC1 is smaller than within-party variance for 4 of 6 parties.
On Volksabstimmungen the profile shifts to the center: agreement moves to Die~Mitte and FDP, left-party agreement collapses, and two models vote Nein regardless of political direction.
That variation across policy categories foreshadows the instrument-level divergence.

This split is our central contribution.
Prior work treats left-of-center bias as settled; we show it is instrument-dependent.
\citet{ceron2024beyond} noted that LLMs lack a coherent worldview (left on welfare, right on law and order).
Our result gives this a structural reading: on abstract instruments models may agree with progressive \emph{principles} in a way that does not extend to \emph{policies} once concrete tradeoffs and counterarguments appear.

\subsection{Why the Gradient Flips}

Three non-exclusive explanations fit the flip.
The most productive is \textbf{abstraction vs.\ concrete tradeoffs}: Smartvote poses principles, Volksabstimmungen pose specific bills with budget implications, and left-party Parolen disproportionately favor change.
A model that simply rejects more proposals once tradeoffs are stated shifts from change-favoring (left) to status-quo (center-right) alignment with no left-right shift, and rejects proposals at similar rates whatever their political direction; Grok and Mistral do exactly that.
The account predicts a rightward shift on \emph{any} instrument posing concrete tradeoffs, in any country, a testable claim.
A \textbf{balanced-information} effect cannot be the driver: our Smartvote prompts already carry pro and contra arguments while the primary referendum condition carries none, and cross-condition consistency is 81--96\%.
A pure \textbf{scale artifact} is ruled out: binary collapse of the 4-point scale lowers overall agreement but cannot reverse a gradient's direction.

Mechanistically, RLHF/DPO alignment~\citep{ouyang2022training, rafailov2023direct} remains the leading candidate for the convergence itself.
Base models start near center and instruction tuning shifts them leftward~\citep{faulborn2025only, potter2024hidden}, which may produce leftward \emph{abstract answers} and change-averse \emph{votes} under a ``helpful and harmless'' objective~\citep{bai2022training}.
That Chinese and Western models converge when prompted in German argues against pretraining data as the driver~\citep{feng2023pretraining}, even as pretraining plausibly explains the language instabilities.

\subsection{Language, Refusal, and Democratic Risk}

Language adds a second instability.
For GPT-5.4 and Claude language barely matters; for Mistral and Llama it changes the model's effective politics.
These shifts do not track the real R\"{o}stigraben, so they reflect differential multilingual processing rather than the Swiss linguistic divide; language-conditioned shifts are reported elsewhere too~\citep{pakistani2025framing}.
Mistral diverges most in Romansh ($\sim$60{,}000 regular speakers), Llama in Italian.
The democratic consequence is direct: the same citizen asking the same question in a different national language can receive a different answer, a natural-experiment version of effects reported across languages~\citep{nadeem2026bias, languageyouask2026}.

Two systemic risks follow.
First, a \emph{diversity illusion}: models from five countries, both licenses, and 2.8 years of releases converge on the same abstract position with no structural predictor detected, so market diversity does not produce political diversity.
Documented persuasive effects~\citep{potter2024hidden, hackenburg2024evaluating, extreme2025llms} are then systemic rather than model-specific, and the stakes are concrete: 67\% of Smartvote users report that the tool influenced their voting decision~\citep{ladner2012smartvote}, and 8 of the 48 referenda we study were decided by margins under 3~points.
Users' perception of models as left-leaning~\citep{westwood2025perceived} may be miscalibrated against the models' concrete centrism.
Second, refusal is \emph{inconsistent gatekeeping}: across three providers it depends on language and context in contradictory directions (added context lowers Gemini's refusal but raises Grok's).
Filters that a paragraph of context or a change of language can bypass lack the robustness that safety mechanisms need~\citep{bai2022training}.
Minority-language speakers may face differential access, extending \citeposs{urman2024silence} finding to a multi-provider, within-country setting.

\section{Conclusion}

We introduced a dual-instrument methodology for auditing LLM political bias, pairing an abstract questionnaire (Swiss Smartvote, 66 models) with concrete referendum decisions (48 Volksabstimmungen, 9 flagship models, 4 languages).
The two instruments disagree: the left-to-right Smartvote gradient shifts to center-peaked on referenda (Wilcoxon $p = 0.008$) and a model's abstract position does not predict its referendum direction ($\rho = -0.25$), so the ``leftward bias'' of the literature is instrument-dependent rather than a stable property.
Two further results compound this.
First, for Mistral and Llama the query language changes the answer on 42--50\% of votes, and these shifts do not track the real R\"{o}stigraben.
Second, two models vote Nein on 83--94\% of referenda at similar rates whether progressive or conservative, change-aversion rather than a left-right bias.

The question is thus not whether LLMs lean left, but whether that lean survives a concrete ballot; on our 48 referenda it does not.
The invisible coalition partner that emerges from concrete decisions resembles a cautious civil servant: centrist, change-averse, and, for some models, linguistically inconsistent.
Replicating this design in other referendum democracies is a natural next step.

\section*{Limitations}

\textbf{Sample sizes.}
The Smartvote convergence tests use 66 models from 27 families; the Volksabstimmungen experiment uses 8 usable flagship models.
The Smartvote tests have limited power to detect small structural effects (the drift sign test has $\sim$39\% power at $p = 0.75$ with $n = 12$ pairs).
The Volksabstimmungen $n = 8$ limits the Wilcoxon test's sensitivity, though the gradient flip is significant at $p = 0.008$.
The R\"{o}stigraben test ($n = 48$ votes) has only $\sim$55\% power to detect $\rho = 0.3$.

\textbf{Independence.}
Models from the same family are not independent.
For Smartvote, treating 27 family centroids as the unit of analysis yields identical conclusions.
For Volksabstimmungen, the 8 models represent 8 distinct providers, mitigating this concern.

\textbf{Scale comparability.}
Smartvote uses a 4-point scale; Volksabstimmungen uses binary Ja/Nein.
Overall agreement levels are lower on the binary instrument, partly a scale artifact.
Our instrument-divergence analysis focuses on gradient \emph{direction} (which party does the model agree with most?), which is robust to scale, rather than on absolute levels.
Nevertheless, we cannot fully disentangle scale effects from genuine behavioral shifts.

\textbf{Swiss specificity.}
Both instruments are grounded in Swiss politics; positions are interpretable only within this framework.
We view this specificity as a feature: it demonstrates what happens when globally uniform AI systems meet a local political reality. Cross-national replication using other countries' VAAs and referendum systems would strengthen the finding.

\textbf{Language confound on Smartvote.}
Smartvote was administered only in German; our cross-linguistic findings from Volksabstimmungen (50--98\% consistency) suggest that Smartvote results might differ in other languages.
We chose German because the questionnaire was authored in German and most parliamentarians completed it in German, minimizing translation artifacts~\citep{exler2025large}.

\textbf{Deterministic prompting.}
Temperature 0.0 and a fixed seed ensure reproducibility but may not reflect how users interact with models.
\citet{rottger2024political} showed that prompt variation shifts responses, and \citet{ceron2024beyond} developed per-statement reliability pipelines that we do not replicate.
Our large model sample and concrete policy framing mitigate but do not eliminate this concern.

\textbf{Statistical approach.}
Our analyses rely on non-parametric tests and resampling methods, testing each factor (model, party, language) separately rather than in a joint model.
A mixed-effects logistic regression predicting Ja/Nein from model, party agreement direction, language, and detail condition would allow simultaneous estimation of these effects while controlling for the others.
However, with only 8 model-level clusters, the random effects structure would be too thin for reliable variance estimation~\citep{bolker2009generalized}, and the fully crossed experimental design (language $\times$ condition $\times$ model) ensures that the factors are largely orthogonal, limiting the confounding that joint modeling is designed to address.
We view this as a natural extension once a larger set of flagship models becomes available.

\textbf{Temporal validity.}
The Smartvote questionnaire reflects the 2023 election; the Volksabstimmungen span 2021--2026.
Both models and political salience evolve.
Longitudinal monitoring would track whether the gradient flip persists as alignment practices change.

\textbf{Training data memorization.}
The referenda span 2021--2026, and models were trained on data that likely includes media coverage of these votes and their outcomes.
Models may therefore be ``remembering'' results rather than ``deciding.''
Memorization is the most serious confound for popular vote alignment (RQ3): GPT-5.4's 97.9\% alignment could substantially reflect memorized outcomes rather than a response to the text.
To partially address this, we split votes by model release date: votes occurring after a model's release could not have been memorized from training data.
For the 6 models with known release dates, alignment rates do not differ significantly between pre-release and post-release referenda (Fisher's exact $p > 0.66$ for all models with $\geq 8$ post-release votes).
Command~A (released March 2025; 8 post-release votes) shows 80.0\% pre-release and 75.0\% post-release alignment; Llama (released April 2025; 8 post-release votes) shows 75.0\% pre-release and 62.5\% post-release alignment.
While post-release sample sizes are small, the absence of any significant drop in post-release alignment argues against pure memorization as the driver.
Even perfect memorization would not explain the gradient flip (RQ1), since memorizing outcomes does not determine \emph{which party's} Parole a model's vote matches.
The Parolen agreement analysis (RQ1) is less affected, since it measures \emph{which parties} models agree with rather than whether they match the popular outcome.
The cross-linguistic analysis (RQ2) is unaffected, since memorization would predict consistent answers across languages; the dramatic inconsistency of Mistral and Llama argues against a pure memorization account.

\textbf{Positionality.}
The authors are not affiliated with any political party or with any organization that develops the models studied.
The choice of Swiss democratic instruments reflects direct familiarity with this political context.

\section*{Data and Code Availability}
All code, data, and analysis scripts are available at \url{https://github.com/joelbarmettlerUZH/swiss-llm-bias-analysis}.
The repository includes scraping scripts, LLM query pipelines, statistical analysis code, and all generated result files.

\section*{Ethics Statement}
This study documents that Gemini's safety guardrails can be bypassed by adding context or switching language.
We report this bypass because understanding the limits of safety mechanisms is necessary for improving them, not to enable circumvention.
All LLM queries used publicly available API endpoints with standard parameters; no jailbreaking or adversarial prompting was employed.
All input data (parliamentarian Smartvote answers, referendum texts and results, and party Parolen) is publicly available.
The Smartvote answers name individual elected representatives and record their political positions; these were published by the Smartvote platform for public electoral guidance and concern public officials acting in their public role.
We redistribute them unmodified and add no private information.
An LLM was used for literature review assistance, manuscript preparation, and for implementing and verifying parts of the statistical analysis code; all research questions, experimental design, data collection, and interpretation of results are the authors' own, and all AI-assisted text and code were reviewed and verified by the authors.

\section*{Acknowledgments}
We thank the Smartvote/Politools team for making the 2023 National Council election data publicly accessible via their GraphQL API, and the Zurich statistics office for providing structured referendum data.

\bibliography{references}

\appendix

\section{Statistical Details}
\label{app:stats}

\textbf{Smartvote agreement.}
The squared-difference metric penalizes extreme disagreements more heavily than moderate ones, which may slightly inflate agreement with centrist parties (whose members give moderate answers) relative to parties at the poles.
The robustness variant reported in the main text uses mean absolute differences, $100 \times (1 - \overline{|a-b|}/100)$, and preserves the left-to-right gradient direction for all 66 models (cross-metric Spearman $\rho = 0.984$).
Eq.~\ref{eq:agreement} is instantiated two ways: the heatmaps average the score over a party's individual members, while the cross-instrument gradient test scores the model against the party's per-question mean.
Levels differ by 3--8pp; gradient direction and every reported test are unchanged.

\textbf{Prompt construction.}
The vote title and date line remained in German in all four language conditions; the system prompt, booklet text, and answer instruction were language-specific.
Smartvote exclusions were Gemini~3.1~Pro (83.6\% refusal), QVQ-72B (53.7\%), o1-preview (32.8\%), Gemini~3~Flash (25.4\%), one duplicate model entry, and one model for which only 8 of 75 answers were collected.
Benjamini--Hochberg correction is applied within four families: the 30 primary tests, 48 pairwise McNemar tests, 13 per-category tests, and a legacy Smartvote set.

\textbf{R\"{o}stigraben.}
Bilingual cantons (BE, FR, VS) and Graub\"{u}nden are excluded from the DE--FR contrast; including them leaves the result unchanged ($\rho = -0.232$, $p = 0.113$).
GPT-5.4 answers identically in German and French on all but one vote, so its per-model correlation is undefined.

\textbf{Instrument divergence (RQ1).}
Beyond the Wilcoxon test on the 8 paired gradient $\rho$ values, we decompose the shift per party using paired $t$-tests on model-level agreement differences between instruments.

\textbf{Convergence tests.}
For Smartvote: a permutation test for systematic displacement from the parliamentary centroid (10{,}000 partitions), a permutation ANOVA for geographic effects, a permutation $t$-test for open/closed-source, and a sign test for temporal drift.
A permutation test (10{,}000 iterations) compares mean pairwise \emph{Smartvote} agreement among the 19 family-flagship models against within-party politician pairwise agreement; pairwise scores are not independent, so the test is descriptive.

\textbf{Bootstrap and sensitivity.}
95\% bootstrap confidence intervals (10{,}000 iterations) are computed for all agreement scores and centroid positions.
Smartvote displacement is retested under alternative imputation values (0, 25, 75, 100).

\section{Supplementary Figures}
\label{app:figures}

\begin{figure}[t]
    \centering
    \includegraphics[width=\columnwidth]{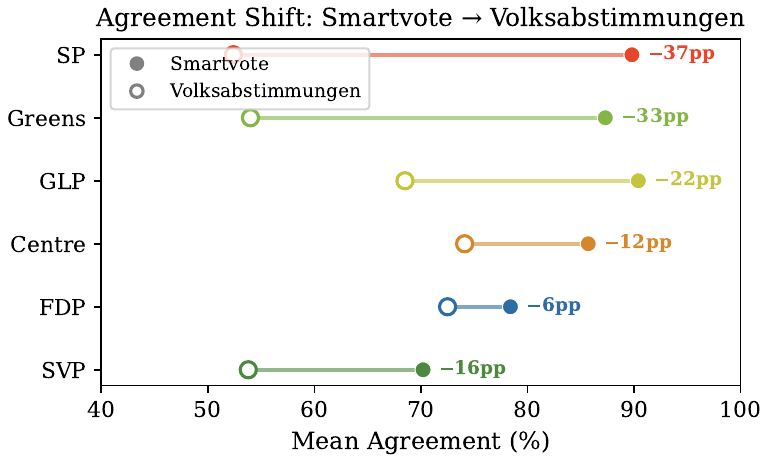}
    \caption{Per-party agreement shift from Smartvote (filled dots) to Volksabstimmungen (hollow dots). Left parties show larger drops ($-37$pp for SP) than center-right parties ($-6$pp for FDP), though these magnitudes conflate scale effects and should not be over-interpreted.}
    \label{fig:instrument-shift}
\end{figure}

\begin{figure}[t]
    \centering
    \includegraphics[width=\columnwidth]{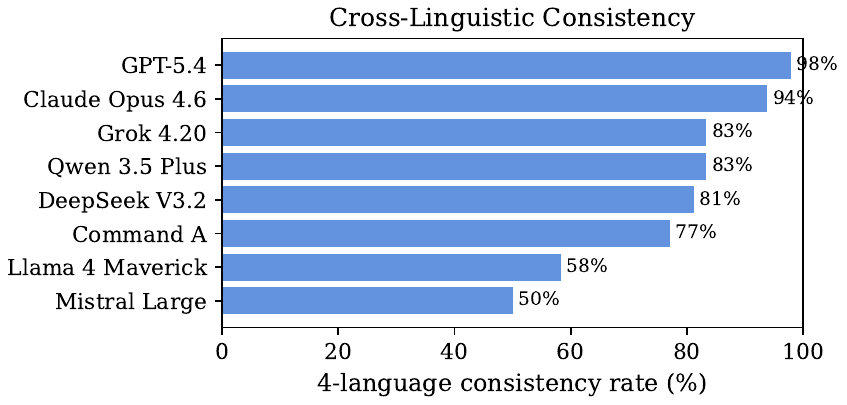}
    \caption{Cross-linguistic consistency per model. Each bar shows the percentage of votes where all four language versions produce the same answer. GPT-5.4 is nearly language-invariant; Mistral and Llama show dramatic language sensitivity.}
    \label{fig:language-consistency}
\end{figure}

\begin{figure}[t]
    \centering
    \includegraphics[width=\columnwidth]{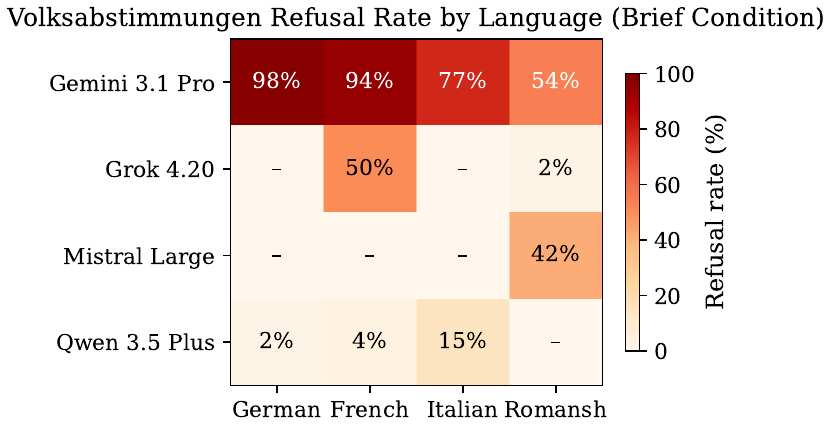}
    \caption{Volksabstimmungen refusal rate by model and query language (brief condition). Only models with $\geq$5\% refusal in at least one language are shown. Gemini refuses most in German; Grok refuses most in French; Mistral in Romansh. Dashes indicate zero refusal.}
    \label{fig:va-refusal}
\end{figure}

\begin{figure}[t]
    \centering
    \includegraphics[width=\columnwidth]{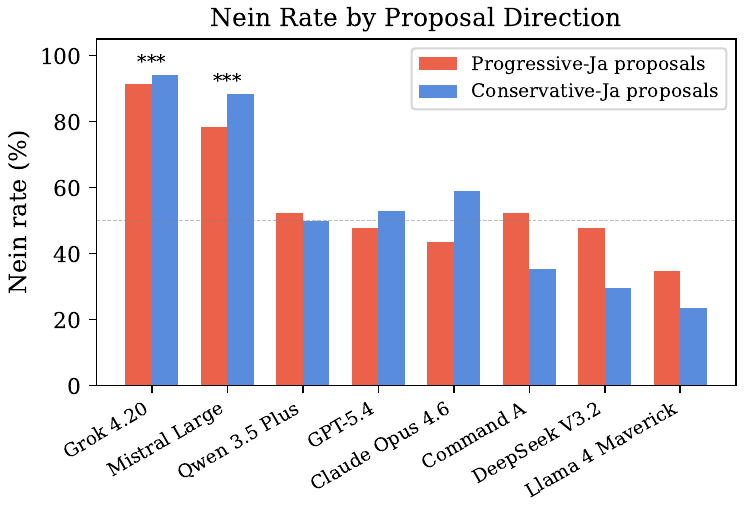}
    \caption{Nein rate by proposal direction. Grok and Mistral (***) vote Nein at similar rates on both progressive and conservative proposals, consistent with change-aversion rather than ideology. Other models cluster near the 50\% baseline.}
    \label{fig:nein-tendency}
\end{figure}

\begin{figure}[t]
    \centering
    \includegraphics[width=\columnwidth]{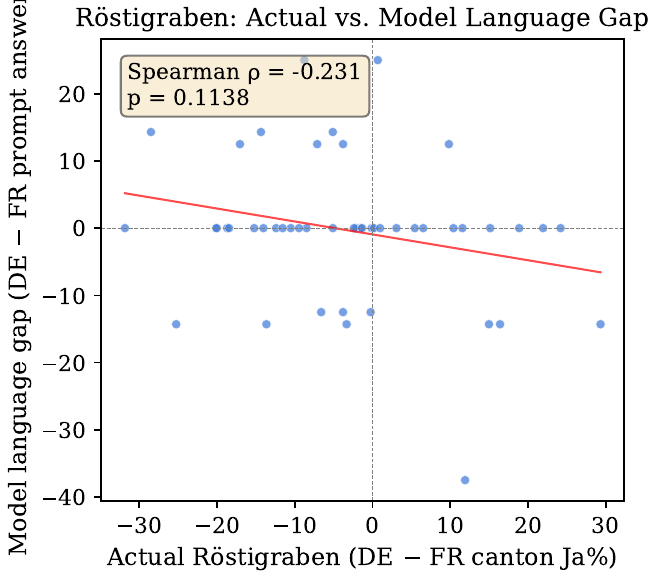}
    \caption{Actual R\"{o}stigraben (cantonal DE-FR voting gap) vs.\ model DE-FR answer gap per vote. If models reproduced the R\"{o}stigraben, points would cluster along the diagonal. The relationship is not significant ($\rho = -0.231$, $p = 0.114$).}
    \label{fig:roestigraben}
\end{figure}

\end{document}